\documentclass[12pt]{article}

\usepackage{graphicx}
\graphicspath{{figures/}} % set a path
\usepackage{caption}
\usepackage{subcaption}
\usepackage{amssymb}
\usepackage{amsmath}
\usepackage{dsfont}				% for \mathbb{1}=\mathds{1}
\usepackage{slashed}			% for slashed symbols
\usepackage[compat=1.0.0]{tikz-feynman}			% for Feynman diagrams 
\usepackage{jheppub}
\newcommand{\mbbone}{\text{\usefont{U}{bbold}{m}{n}1}}
\MakeRobust{\mbbone}

\def\b{\beta}

\def\d{\delta}
\def\D{\Delta}

\def\eps{\varepsilon}
\def\f{\frac}

\def\l{\left}
\def\la{\langle}
\def\ra{\rangle}

\def\mc{\mathcal}
\def\m{\mu}

\def\p{\partial}
\def\r{\right}

\def\be{\begin{equation}}
\def\ee{\end{equation}}
\def\bea{\begin{eqnarray}}
\def\eea{\end{eqnarray}}
\def\ba{\begin{array}}
\def\ea{\end{array}}
\def\bc{\begin{center}}
\def\ec{\end{center}}
\def\bl{\begin{flushleft}}
\def\el{\end{flushleft}}
\def\br{\begin{flushright}}
\def\er{\end{flushright}}
\def\bi{\begin{itemize}}
\def\ei{\end{itemize}}
\def\bt{\begin{tabular}}
\def\et{\end{tabular}}

\def\be{\begin{equation}}
\def\ee{\end{equation}}
\def\bea{\begin{eqnarray}}
\def\eea{\end{eqnarray}}
\def\ba{\begin{array}}
\def\ea{\end{array}}
\def\bc{\begin{center}}
\def\ec{\end{center}}
\def\bl{\begin{flushleft}}
\def\el{\end{flushleft}}
\def\br{\begin{flushright}}
\def\er{\end{flushright}}
\def\bi{\begin{itemize}}
\def\ei{\end{itemize}}
\def\bt{\begin{tabular}}
\def\et{\end{tabular}}
\title{Diagram dependance of the $\mathbb{CP}(1)$ beta function}

\author{Kasra Kiaee,}
\author{Alexander Monin}

\affiliation{University of South Carolina \\
712 Mains St, 404 \\
Columbia SC, 29208

}
\emailAdd{kkiaee@email.sc.edu}
\emailAdd{amonin@mailbox.sc.edu}

\abstract{We revisit the computation of the beta function in the two-dimensional $\mathbb{CP}(1)$ sigma model. We show that in different schemes, different diagrams are responsible for the running, such as momentum-independent tadpoles or even UV-finite bubble diagrams. We also comment on the relation between the beta functions and the energy dependence of scattering amplitudes.}

\notoc
\begin{document}

\maketitle

\newpage

\section{Introduction}

The notion of running coupling is one of the most used and helpful concepts in both high-energy and condensed matter physics. According to the modern perspective provided by Wilson, the running can be viewed, albeit with subtleties (see~\cite{Joyce:2022ydd}), as a sequential integration of hard modes in the path integral. Wilsonian reasoning differs somewhat when considering threshold effects and the generation of new, even non-local, operators compared to the usual RG evolution practiced in high-energy physics. The latter being simply summing large logarithms. However, these differences become irrelevant~\cite{Polchinski:1983gv} at energies much lower than the cutoff of the theory.

Demanding that correlators (or scattering amplitudes) be independent of the scale $\Lambda$ at which the theory is defined leads to a new set of couplings at each scale. It is in this sense that we say the couplings $g_i(\Lambda)$ are scale-dependent. There is another related notion of running: the dependence of a scattering rate on the energy of the process. The two can, in principle, be connected using the Callan-Symanzik equation. However, since the former is scheme-dependent in perturbation theory\footnote{The running is scheme dependent beyond two loops.}, the correspondence is not one-to-one.

The recent paper~\cite{Donoghue:2019clr} presents valid criticism of power-law running coupling constants as applied to the asymptotic safety scenario. It is correctly argued that the asymptotic behavior of the scattering amplitude (or other physical observables) with energy does not necessarily have the same form as the dependence of the coupling on the cutoff $\Lambda$. However, there are a couple of confusing claims that need clarification.  For instance, the argument that tadpole graphs are momentum independent, and that these graphs do not know about the momentum scales of the physical reactions, may be interpreted as suggesting that these graphs do not contribute to the couplings' running and thus should be omitted altogether. 

Similarly, one may get the impression that the usefulness of the beta function, defined as the coupling's running with the cutoff, may be restricted to Euclidean QFT, and that the Wilsonian approach is somehow not applicable to high-energy physics. Although there are indeed subtleties present uniquely in Lorentzian signature, such as soft and collinear divergencies, this is not the point of criticism, which concerns mostly the relationship between the behavior of scattering amplitudes at high energy and the asymptotic behavior of the coupling.

Given the possibility of misinterpretation, we believe that some clarification is in order, if only for pedagogical reasons. Which diagrams contribute to the beta function is scheme-dependent. For instance, it is customary to drop tadpoles when dealing with dimensional regularization in massless theories, however, it is mandatory to keep them if cutoff regularization is used. In what follows, we present the beta-function computation in the $\mathbb{CP}(1)$ sigma model in $1+1$ dimensions. Keeping track of the cutoff dependence is merely a tool to compute the beta function. It is necessary to take into account the tadpole diagrams to reproduce the correct running. We also perform the computation of the beta function using dimensional regularization and show that despite different origins the corresponding beta functions are the same. Moreover, we show that in this case, the beta function is consistent with the energy dependence of the scattering rate.

\section{Scattering amplitude\label{sec:cutoff}}

We consider the $\mathbb{CP}(1)$ sigma model in two dimensions, given by the following Lagrangian\footnote{This Lagrangian is equivalent to the more frequently used one obtained by rescaling the field $\phi_0$.}
\be
\mc L = \f{\p \phi_0 \p \bar \phi_0}{\l ( 1+ \f{g_0^2}{2} \bar \phi_0 \phi_0 \r )^2},
\ee
where $\phi_0$ is a complex scalar (bare) field and $g^2_0$ is a bare coupling. Expanding the denominator in powers of $g^2_0$ and integrating by parts, we obtain an equivalent form for the Lagrangian
\be
\label{eq:expandedlagrangian}
\mc L = \p \phi_0 \p \bar \phi_0 + \f{g_0^2}{4} \, \bar \phi_0^2 \p^2 \phi_0^2 - \f{g_0^4}{12} \, \bar \phi_0^3 \p^2 \phi_0^3 + \dots ,
\ee
providing four-, six-, and higher-leg interaction vertices.

The scattering amplitude can be found by using the LSZ procedure. Thus, all the intermediate computations can be done in Euclidean space and continued to Lorentzian signature when needed. One thing to note is that diagrams in this theory are not only UV but also IR divergent. To regulate the latter, we introduce a small mass $m^2$. 

%We would like to illustrate how the dependence of the scattering amplitude on the energy can be determined by tracing the dependence of the bare coupling on the cutoff $\Lambda$.

\begin{figure}[h]
    \centering
     
    \begin{subfigure}[b]{0.3\textwidth}
            
            \centering
             \begin{tikzpicture}[baseline=(x)]

                \begin{feynman}

                    \vertex(x);
                    \vertex[above=0.6cm of x,dot](a){};
                    \vertex[left=1cm of a] (b);
                    \vertex[right=1cm of a] (c);
                    \vertex[above=1cm of a] (d);

                    \diagram*{
                       (b)--[very thick,fermion](a)--[very thick,fermion,out=170,in=180,looseness=1.5](d)
                       --[very thick,fermion,out=0,in=10,looseness=1.5](a)--[very thick,fermion](c);
                    };
                \end{feynman} 
            \end{tikzpicture}     
            \caption{\label{fig:2ptFunction}}
    \end{subfigure}
    \hfil
     \begin{subfigure}[b]{0.3\textwidth}
            
            \centering
            \begin{tikzpicture}[baseline=(a)]

                \begin{feynman}

                    \vertex[dot](a){};
                    \vertex[left=1cm of a] (b);
                    \vertex[right=1cm of a] (c);
                    \vertex[above=1cm of a] (d);
                    \vertex[below left=1cm of a] (e);
                    \vertex[below right=1cm of a] (f);

                    \diagram*{
                       (b)--[very thick,fermion](a)--[very thick,fermion,out=170,in=180,looseness=1.5](d)
                       --[very thick,fermion,out=0,in=10,looseness=1.5](a)--[very thick,fermion](c);
                       (e)--[very thick,fermion](a)--[very thick,fermion](f)
                    };
                \end{feynman} 
            \end{tikzpicture}        
            \caption{\label{fig:4ptFunction}}
    \end{subfigure}
   \caption{\label{fig:Tadpoles}  Diagrams that include tadpole integrals. Diagram \ref{fig:2ptFunction} shows up in computing two point function, while we face the \ref{fig:4ptFunction} diagram in calculation of four point functions.}     
\end{figure}
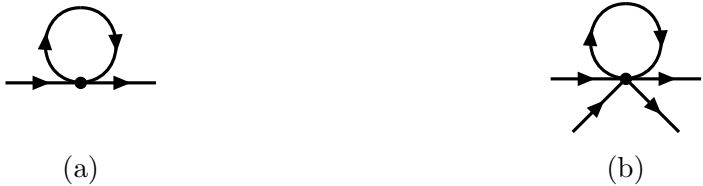

At one loop, the two-point function receives a contribution from the diagram depicted in Figure~\ref{fig:2ptFunction}
\be
\mathcal{I}_{2}=g_0^2\int\frac{d^2 k}{(2\pi)^2}\frac{(k+p)^2}{q^2+m^2}.
\ee
Neglecting terms that vanish in the limit $\Lambda \to \infty$, produces
\be
\mathcal{I}_2=\frac{g_0^2}{4\pi}\Lambda^2+\frac{g_0^2}{4\pi}(p^2-m^2)\log{\frac{\Lambda^2}{m^2}},
\ee
which corresponds to the following wave function renormalization
\be
 \label{eq:fieldcutoff}
\la \tilde \phi_0(p) \tilde {\bar \phi}_0(-p)\ra = \f{Z}{p^2},
\ee 
with
\be
\label{eq:WFRenorm}
 Z = 1+\f{g_0^2}{4\pi} \log\f{\Lambda^2}{m^2}.
\ee
There are also corrections to the bare mass parameter $m^2$, which we neglect since we are focusing on the running of the coupling constant $g^2_0$.

\begin{figure}[h]
    \centering
     
    \begin{subfigure}[b]{0.3\textwidth}
            
            \centering
            \begin{tikzpicture}[baseline=(x)]

                \begin{feynman}

                    \vertex(x);
                    \vertex [above=1.4cm of x,dot](a){};
                    \vertex[above left=1cm of a] (b){$p_1$};
                    \vertex[below left=1cm of a] (c){$p_2$};
                    \vertex[right=1.3cm of a,dot] (d){};
                    \vertex[above right=1cm of d] (e){$p_3$};
                    \vertex[below right=1cm of d] (f){$p_4$};

                    \diagram*{
                       (b)--[very thick,fermion](a)--[very thick,anti fermion](c);
                       (e)--[very thick,anti fermion](d)--[very thick,fermion](f);
                       (a)--[out=70,in=110,very thick,looseness=1.3,fermion](d)
                       --[out=-110,in=-70,very thick,looseness=1.3,anti fermion](a)  
                    };
                \end{feynman} 
            \end{tikzpicture}         
            \caption{\label{fig:sChannel}}
    \end{subfigure}
    \hfil
     \begin{subfigure}[b]{0.3\textwidth}
            
            \centering
            \begin{tikzpicture}[baseline=(a)]

                \begin{feynman}

                    \vertex (a){};
                    \vertex[above=0.57cm of a,dot] (b){};
                    \vertex[below=0.57cm of a,dot] (c){};
                    \vertex[above left=1cm of b] (d){$p_1$};
                    \vertex[above right=1cm of b] (e){$p_3$};
                    \vertex[below right=1cm of c] (g){$p_4$};
                    \vertex[below left=1cm of c] (f){$p_2$};

                    \diagram*{
                       (d)--[very thick,fermion](b)--[very thick,fermion](e);
                       (f)--[very thick,fermion](c)--[very thick,fermion](g);
                        (b)--[out=-20,in=20,very thick,looseness=1.5,fermion](c)--[out=-200,in=-160,very thick,looseness=1.5,fermion](b);  
                    };
                \end{feynman} 
            \end{tikzpicture}         
            \caption{\label{fig:tChannel}}
    \end{subfigure}
     \hfil
    \begin{subfigure}[b]{0.3\textwidth}
            
            \centering
            \begin{tikzpicture}[baseline=(a)]

                \begin{feynman}

                    \vertex (a);
                    \vertex[right=1.1cm of a] (a1);
                    \vertex[above=0.5cm of a,dot] (b){};
                    \vertex[below=0.5cm of a,dot] (c){};
                    \vertex[above left=1cm of b] (d){$p_1$};
                    \vertex[above right=1cm of a1] (e){$p_3$};
                    \vertex[below right=1cm of a1](g){$p_4$};
                    \vertex[below left=1cm of c] (f){$p_2$};

                    \diagram*{
                       (d)--[very thick,fermion](b)--[very thick,fermion, out=5,in=135](a1);
                       (f)--[very thick,fermion](c)--[very thick,fermion, out=-5,in=-135](a1);
                       (e)--[very thick,anti fermion](a1)--[very thick,fermion](g);
                       (b)--[out=-20,in=20,very thick,looseness=1.5,fermion](c)--[out=-200,in=-160,very thick,looseness=1.5,fermion](b)  
                    };
                \end{feynman} 
            \end{tikzpicture}         
            \caption{\label{fig:uChannel}}
    \end{subfigure}
     
\caption{\label{eq:BubbleSTU}  The diagrams, which include the bubble integrals, contributing to four point function are shown. The S-channel \ref{fig:sChannel} diagram is UV-finite but IR-divergent, while \ref{fig:tChannel} and \ref{fig:uChannel}  are UV-divergent. }     
\end{figure}
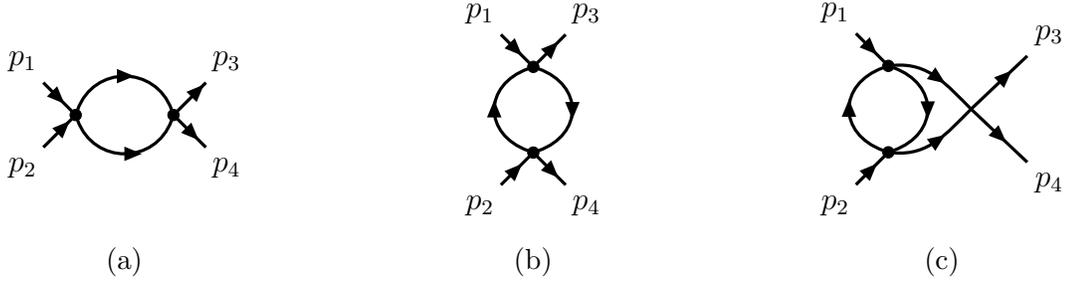

Moving on to computing the (amputated) four-point function, we observe that there are several contributions at one loop. Namely, those depicted in Figure~\ref{fig:4ptFunction} and Figure~\ref{eq:BubbleSTU}. We will refer to the diagrams in Figure~\ref{eq:BubbleSTU} as $s$, $t$, and $u$-channel diagrams, even though there are no different channels in two dimensions. For the $s$-channel the diagram is not UV divergent. However, it is instructive to show that its contribution is necessary to make the scattering amplitude IR finite. The corresponding integral is given by
\be 
\mc A_{s} = \f{g_0^4}{2} (p_1+p_2)^4\int\frac{d^2 k}{(2\pi)^{2}}\int^{1}_{0}dx \frac{1}{(k^2+m^2)((p_1+p_2-k)^2+m^2)}\ee
Introducing the Feynman parameter and performing the momentum integration, we arrive at
\be
\mc A_{s} =\f{g_0^4}{8 \pi} (p_1+p_2)^4 \int_0^1 \f{dx}{m^2+(p_1+p_2)^2 \, x(1-x)}.
\ee
This integral can be done exactly, however, the leading IR divergent term in the limit $m\to 0$ can be easily extracted by cutting off the integral at $x_1=m^2/(p_1+p_2)^2$ and $x_2=1-m^2/(p_1+p_2)^2$, leading to
\be
\int_0^1 \f{dx}{m^2+(p_1+p_2)^2 \, x(1-x)} \approx \f{2}{(p_1+p_2)^2} \log\f{(p_1+p_2)^2}{m^2},
\ee
which results in
\be
\label{eq:4pointSchannel}
\mc A_s = \f{g_0^2}{4\pi}(p_1+p_2)^2 \log\f{(p_1+p_2)^2}{m^2}.
\ee

The $t$ channel diagram is only marginally more involved. The corresponding integral has the following form
\be 
\mc A_t=g^{4}\int\frac{d^2 k}{(2\pi)^2} \frac{(k+p_1)^2 (k+p_3)^2}{(k^2+m^2)((k+p_1-p_3)^2+m^2)}.
\ee
After a bit of algebra we obtain
\be
\mc A_t=\f{g_0^4}{4\pi}\Lambda^2+\f{g_0^4}{4\pi} \int dx \l [ \l ( (q_{13}+q_{42})^2-2\D_{13} \r ) \log\f{\Lambda^2}{\D_{13}} + \f{q_{13}^2q_{42}^2}{\D_{13}} + \D_{13} - (q_{13}+q_{42})^2\r ],
\ee
where the following notations have been introduced
\be
q_{ij} = p_i (1-x)+p_j x,
\ee
and
\be
\D_{ij} = m^2+x(1-x)(p_i-p_j)^2.
\ee
The divergent part can be easily extracted, yielding
\be
\label{eq:4pointTchannel}
\mc A_{t} = \f{g_0^4}{4\pi}\Lambda^2 + \f{g_0^4}{4\pi}(p_1+p_2)^2 \log\f{\Lambda^2}{(p_1-p_3)^2}.
\ee
To analyze the IR structure, we note that only the term $\f{q_{13}^2q_{42}^2}{\D_{13}}$ is potentially IR divergent. Performing the analytic continuation to Minkowski space and setting external momenta on-shell $p_i^2=0$, we observe that this term is finite in the limit $m^2\to 0$
\be
\f{q_{13}^2q_{42}^2}{\D_{13}} \underset{m^2\to0}{=} -2 (p_1p_3) x (1-x).
\ee 
The $u$-channel can be obtained from the result for the $t$-channel by interchanging $p_3\leftrightarrow p_4$
\be
\label{eq:4pointUchannel}
\mc A_{u} = \f{g_0^4}{4\pi}\Lambda^2 + \f{g_0^4}{4\pi}(p_1+p_2)^2 \log\f{\Lambda^2}{(p_1-p_4)^2}.
\ee
Lastly, the tadpole diagram in Figure~\ref{fig:4ptFunction} produces
\be
\mc A_{6} = -3\f{g_0^2}{4\pi}(p_1+p_2)^2 \log\f{\Lambda^2}{m^2}.
\ee

Collecting all terms together, we get the following expression for the scattering amplitude
\begin{align}
\label{eq:amplitude4pointcutoff}
\mc M &= Z^2 (g_0^2 s+\mc A_s +\mc A_t +\mc A_u +\mc A_6)=g_0^2 s \l [ 1 - \f{g_0^2}{4\pi} \log \f{\Lambda^2}{-s} + \f{g_0^2}{4\pi} \log \f{\Lambda^2}{-t} + \f{g_0^2}{4\pi} \log \f{\Lambda^2}{-u} \r ],
\end{align}
with $s$, $t$, $u$ being the standard Mandelstam variables. The corresponding beta function can be found by differentiating $g_0^2$ with respect to $\log \Lambda$, resulting in
\be
\label{eq:betafunction}
\b(g^2) = -\f{g^4}{2\pi},
\ee
which is the usual result for the beta function of the $\mathbb{CP}(1)$ sigma model~\cite{Polyakov:1975rr}. Energy dependence of the scattering amplitude obviously can be extracted from the running of the coupling dictated by so obtained beta function.  

A comment is in order. The contribution from the $t$ and $u$-channels, \eqref{eq:4pointTchannel} and \eqref{eq:4pointUchannel}, also contain momentum-independent quadratic UV divergences. We neglected these since they can be renormalized by an operator of the form $\bar\phi^2\phi^2$, thus not contributing to the running of~$g^2$.

%\subsection{Dimensional regularization \label{sec:DimReg}}

For completeness, we present the result for the beta function computed using dimensional regularization. The main advantage of this scheme is that the number of space-time dimensions serves as a regulator for both UV and IR divergences. As a consequence, tadpole diagrams are automatically zero and can be neglected. The downside is that, if not careful enough, $1/\eps$ poles from IR divergences may lead do the wrong choice of counterterms. 

We will not worry about this subtlety because, as illustrated above using cutoff regularization, the scattering amplitude is IR finite. Therefore, no $1/\eps$ pole coming from IR divergences can contribute to the final expression. Thus, the only loop diagrams we need to take into account are those in Figure~\ref{eq:BubbleSTU}. Introducing, as customary for dimensional regularization, the renormalized coupling
\be
g_0^2=\m^\eps g^2 \l ( 1+ \d_{g^2}\r ),
\ee
we obtain the following expressions (as before, omitting finite non-$\log$-enhanced terms)
\begin{align}
\label{eq:4pointDimReg}
\mc A_s & = 
-\f{\m^\eps g^4}{4\pi}(p_1+p_2)^2 \l (\f{2}{\eps} + \log\f{\m^2}{(p_1+p_2)^2} \r ),\\
\mc A_{t} &= \f{\m^\eps g^4}{4\pi}(p_1+p_2)^2 \l (\f{2}{\eps} + \log\f{\m^2}{(p_1-p_3)^2} \r ),\\
\mc A_{u} &=\f{\m^\eps g^4}{4\pi}(p_1+p_2)^2 \l (\f{2}{\eps} + \log\f{\m^2}{(p_1-p_4)^2} \r ).
\end{align}
Taking into account the diagram corresponding to the counterterm $\d_{g^2}$, we find the following expression for the scattering amplitude
\be
\mc M =\m^\eps g^2 s \l [1 + \d_{g^2} + \f{g^2}{4\pi} \l (\f{2}{\eps} - \log\f{\m^2}{-s} + \log\f{\m^2}{-t} + \log\f{\m^2}{-u} \r ) \r ].
\ee
As a result,
\be
\d_{g^2} = - \f{g^2}{4\pi} \f{2}{\eps},
\ee
which can be used in\footnote{This expression is valid only at one loop. In general, the beta function is given by
\be
\b = g^4 \f{\p}{\p g^2} z_1, 
\ee
where $z_1$ corresponds to a simple pole in the expansion
\be
\log (1 + \d_{g^2} ) = \f{z_1}{\eps}+ \f{z_2}{\eps^2} + \dots
\ee}
\be
\b = g^4 \f{\p }{\p g^2} (\eps \d_{g^2}).
\ee
to reproduce the same beta function \eqref{eq:betafunction}.

\section{Discussion and conclusion \label{sec:Discussion}}

Hopefully, the presented simple computations show that whether tadpole diagrams contribute to the beta function is scheme dependent. In dimensional regularization, they are automatically zero and thus can be neglected. However, in cutoff regularization, they are necessary to reproduce the correct result. Even though tadpole diagrams are momentum independent, in the latter case, the running of the coupling is obtained from the cutoff dependence. 

The question now arises: what good is the beta function, defined in this way, for? The only interest in knowing the beta function lies in its relationship with physically relevant information, if it can be related to physical observables such as  scattering rates or scattering amplitudes. In particular, it is interesting to understand the asymptotic dependence of these observables at high energies. Formally, any (scheme-dependent) expression for the beta function can be used to extract information about scheme-independent observables via the Callan-Symanzik equation. However, in practice, we are usually limited by perturbation theory making it problematic to extrapolate the scattering amplitude to arbitrary high energies.

At the same time, a simple argument presented by Weinberg in~\cite{Weinberg:1980gg} shows that the energy dependence of scattering rates can be obtained from the corresponding behavior of coupling constants. Indeed, given a scattering rate, its energy dependence is fixed by dimensional analysis to have the following form
\be
\mc R(E) = \m^\D f\l ( \f{E}{\m}, X, g_i(\m) \r ),
\ee
where $\m$ is the renormalization scale, $g_i(\m)$ are dimensionless couplings, and $X$ represents all other dimensionless parameters. Since the rate should be $\m$-independent, we may choose to renormalize at $\m = E$, leading to
\be
\mc R(E) = E^\D f \l ( 1, X, g_i(E) \r ),
\ee
therefore, the asymptotic behavior of the rate is defined by that of the coupling constants.

A couple of comments are in order. First, the result for the scattering rate is only as reliable as our control of the coupling constants. If the latter grow sufficiently at high energy, there is no hope of making a definite conclusion about the behavior of the scattering rate. Second, if there are additional mass parameters $m$ one has to make sure that the limit $m/\m \to 0$ is well-defined. Thus, if the coupling constants reach a fixed point\footnote{For subtleties related to the vanishing of the beta function and the trace of the energy momentum tensor see \cite{Luty:2012ww}.}, as considered by Weinberg in~\cite{Weinberg:1980gg}, and the scattering rate has a smooth behavior in the massless limit, it exhibits a simple power-like energy scaling. The presence of additional mass parameters also opens the possibility of redefining the couplings, thereby changing their asymptotic behavior. Below, we demonstrate two examples illustrating these subtleties.

\paragraph{Example 1}

Consider a theory with a marginal coupling $\lambda$ and a mass scale $m$. A typical expression for the scattering amplitude (at one loop) renormalized at a scale $\m$ is
\be
\mc M(E) = \lambda(\m)+b \lambda^2(\m) \log \f{E^2+m^2}{\m^2} \equiv 
\lambda(\m)+b \lambda^2(\m) \log \l [ \f{E^2}{\m^2} +g_2(\m) \r ].
\label{eq:1loopAmplitude}
\ee
From the renormalization scale independence
\be
\f{d}{d\log \m} \mc M(E) = 0,
\ee
it follows that\footnote{Note that the coupling defined in this way always runs, as opposed to the usual running only above the scale $m$. This is a manifestation of the fact that the coupling does not have the same functional dependence of the scale as the scattering amplitude.}
\be
\b_\lambda \equiv \dot \lambda =2b\lambda^2,
\ee
whose solution is
\be
\lambda(\m) = \f{\lambda_0}{1-b \lambda_0\log {\m^2}/{\m_0^2}},
\ee
where $\lambda_0$ and $\m_0$ are constants, $\lambda(\m_0) = \lambda_0$. The running of $g_2$ is trivial
\be
\b_{g} \equiv \dot g_2 = -2 g_2.
\ee
As a result, the amplitude can be found by solving the Callan-Symanzik equation 
\be
\label{eq:Example1Amplitude}
\f{\p}{\p\log E} \mc M = \b_\lambda \f{\p}{\p \lambda} \mc M + \b_{g}\f{\p}{\p g_2} \mc M,
\ee
leading to
\be
\mc M(E) =\lambda(E)+b \lambda^2(E) \log \l [ 1+ g(E) \r ]= \lambda(E)+b \lambda^2(E) \log \l ( 1+ \f{m^2}{E^2}\r )
\label{eq:Amplitude1}
\ee
which up to $\lambda_0^2$ terms, indeed coincides with \eqref{eq:1loopAmplitude}. We see explicitly that $\mc M(E) \neq \lambda (E)$. The equality can be reached only for $E \gg m$. At the same time, perturbation theory breaks down for sufficiently large energies, $E>E_{L}$, with $E_L$ defined from
\be
b \lambda_0 \log \f{E_L^2}{\m^2}=1.
\ee

\paragraph{Example 2}

The reason no singularities appear in \eqref{eq:Example1Amplitude} for $m \ll E < E_L$ is that the amplitude is regular in the limit $m\to 0$. There are situations, for instance, discussed in~\cite{Donoghue:2023yjt, Buccio:2023lzo}, where this is not the case, at least formally. Imagine that in \eqref{eq:1loopAmplitude} we redefined the coupling according to
\be
\lambda = \tilde \lambda + a \tilde \lambda^2 \log \f{m^2}{\m^2}.
\label{eq:ChangeCouplings}
\ee
The amplitude's dependence on energy would not be affected, though its form as a function of $\tilde \lambda$ would be different
\be
\mc M(E) = \tilde \lambda(\m)+b \tilde \lambda^2(\m) \log \f{E^2+m^2}{\m^2}+ a \tilde \lambda^2(\m) \log \f{m^2}{\m^2}
\label{eq:2loopAmplitude}.
\ee
The corresponding beta function becomes
\be
\dot {\tilde \lambda} = 2(a+b) \tilde \lambda^2,
\ee
and the amplitude 
\be
\mc M(E) = \tilde \lambda(E) + b \tilde \lambda^2(E) \log \l ( 1+ \f{m^2}{E^2}\r )- a \tilde \lambda^2(E) \log \f{E^2}{m^2}.
\label{eq:Amplitude2}
\ee
In this case, perturbativity is not broken in a smaller region
\be
\tilde \lambda(E) \log \f{E^2}{m^2} < 1,
\ee
where the two amplitudes \eqref{eq:Amplitude1} and \eqref{eq:Amplitude2} coincide. The reason for the presence of the large $\log E^2/m^2$ term is that the perturbative expression for the amplitude \eqref{eq:1loopAmplitude} does not have a smooth limit $m\to 0$. The same amplitude \eqref{eq:1loopAmplitude}, written in terms of $\lambda$, is IR finite. The singular behavior of the redefinition \eqref{eq:ChangeCouplings} artificially introduces the singularity in the amplitude. In other words, there are different ways of taking the limit $m \to 0$,depending on whether $\lambda (\m)$ or $\tilde \lambda (\m)$ is held fixed, and additional care should be taken in choosing the one with physical relevance.

\paragraph{Example 3}

Lastly, we would like to comment on the effects of irrelevant couplings on the running of marginal ones. Consider a theory with an irrelevant coupling $G$, say of dimension $-2$, and a marginal coupling\footnote{For instance, these could be a gauge coupling and gravity. Corrections of this sort were considered in~\cite{Tang:2008ah}} $g$. The only contribution that could lead to a modification of the marginal coupling has the following form
\be
(G \Lambda^2)^\#.
\ee
Using dimensional regularization, this term cannot appear in the scattering amplitude. Even though it is possible to choose a scheme where we do not eliminate the contribution of the irrelevant coupling to the Wilsonian running of the marginal one, this running does not mean that the momentum dependence of the scattering amplitude suddenly changes. Indeed, a schematic form for the amplitude in this case is
\be
\mc M (E) = g^2(\Lambda) + A \f{g^4(\Lambda)}{16\pi^2} \log \f{E}{\Lambda} + B \f{g(\Lambda)\bar G(\Lambda)}{16\pi^2},
\ee
where $A$ and $B$ are constant coefficients, and we introduced the following dimensionless coupling
\be
\bar G (\Lambda) = \Lambda^2 G.
\ee

It is clear that the only source of the energy dependence comes from the second term, while the running of the marginal coupling is governed not only by the coefficient $A$, but also by the irrelevant coupling $\bar G$.
\be
\label{eq:IrrelevantInduced}
\f{d g^2(\Lambda)}{d\log \Lambda} = A \f{g^4(\Lambda)}{16\pi^2}-B \f{g(\Lambda)}{8\pi^2} \bar G(\Lambda).
\ee
Extending this result to arbitrary large $\Lambda$, where the coupling $\bar G$ becomes large, is dubious. But even if we could do that, it would not mean that the scattering amplitude would scale quadratically with the energy $E^2$ since the scattering amplitude is not given by just $g^2(E)$, but rather by
\be
\mc M (E) = g^2(E) + B \f{g(E)\bar G(E)}{16\pi^2}.
\ee
Taking the running in \eqref{eq:IrrelevantInduced} into account, we can easily convince ourselves (by computing a derivative) that the only energy dependence of the amplitude $\mc M$ still comes from the $A$ term.

\section*{Acknowledgements}

Similar results were obtained in~\cite{Buccio:2024abc} and we thank the authors for fruitful discussions as their calculations have developed. The work is supported in part by the National Science Foundation under Award No. 2310243.

\appendix

\section{Initial and final state radiation \label{app:IRDiv}}

The standard approach to eliminating IR divergences from loop diagrams in theories with massless particles is to additionally consider processes with soft mode radiation. Only the sum of cross sections with and without radiation lead to the IR safe quantities. We have already shown that the $2\to 2$ scattering amplitude is IR finite. For completeness, in this section, we demonstrate that the soft mode radiation processes are also IR finite.

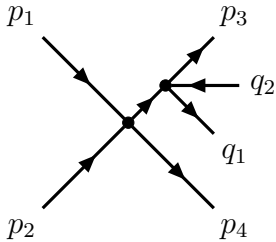
\begin{figure}[h]
    \label{fig:IRdivdiagrams}
    \centering
               
            \centering
            \begin{tikzpicture}[baseline=(x)]

                \begin{feynman}

                    \vertex[dot](a){};
                    \vertex[above left=2cm of a] (b){$p_1$};
                    \vertex[below left=2cm of a] (c){$p_2$};
                    \vertex[above right=0.7cm of a,dot] (d){};
                    \vertex[above right=2cm of a] (e){$p_3$};
                    \vertex[below right=1.3cm of d] (f){$q_1$};
                    \vertex[right=1.3cm of d] (g){$q_2$};
                    \vertex[below right=2cm of a] (h){$p_4$};

                    \diagram*{
                       (b)--[very thick,fermion](a)--[very thick,anti fermion](c);
                       (e)--[very thick,anti fermion](d)--[very thick,fermion](f);
                       (h)--[very thick,anti fermion](a)--[very thick,fermion](d)--[very thick,anti fermion](g);
                    };
                \end{feynman} 
            \end{tikzpicture}  
        
\caption{\label{fig:IR-Div} $2\to4$ scattering diagram that will cancel the $2\to2$ IR divergence. These diagrams are physically involved in the process, as in any experiment, there is a minimum limit energy involved.}     
\end{figure}

At order we are interested in, there are diagrams contributing to $2\to 4$ process like the one presented in Figure~\ref{fig:IR-Div}. The contribution to the scattering amplitude corresponding to this diagram is given by
\be 
\mc M_{3}=g^4 s \frac{ (p_3+q_1)^2}{(p_4+q_1+q_2)^2}
\ee
Parametrizing momenta with energy and unit vectors corresponding to the direction of motion for each particle
\be
p_i = E_i (1,\vec n_i), ~~ q_i = \eps_i (1,\vec m_i),
\ee
and neglecting terms proportional to $\eps_i^2$, we obtain
\be
\mc M_{3}=g^4 s \frac{\eps_1(1-\vec n_3 \vec m_1)}{\eps_1(1-\vec n_3 \vec m_1) + \eps_2(1-\vec n_3 \vec m_2)}.
\ee
Clearly, this expression can be singular only if both terms in the denominator vanish. However, in this case, the numerator also vanishes, making the quantity in question IR finite. The same logic can be used to show that other diagrams corresponding to soft mode radiation are also IR safe.

\bibliographystyle{JHEP}
\bibliography{CP1-beta}{}

\end{document}